\documentclass[10pt,aps,prr,twocolumn,superscriptaddress,showpacs]{revtex4-1}
\usepackage{amsmath}
\usepackage{amsfonts}
\usepackage{graphicx} 
\usepackage{float}
\usepackage{dcolumn}
\usepackage{bm}
\usepackage{color}

\newcommand*\xbar[1]{%
   \hbox{%
     \vbox{%
       \hrule height 0.5pt 
       \kern0.3ex
       \hbox{%
         \kern-0.1em
         \ensuremath{#1}%
         \kern-0.1em
       }%
     }%
   }%
} 

\newcommand{\Deltabar}{{\xbar{\Delta}}}
\newcommand{\varphivec}{\mbox{\boldmath$\varphi$}}
\newcommand{\etavec}{\mbox{\boldmath$\eta$}}

\newcommand\Wtilde{\stackrel{\,\sim}{\smash{W}\rule{0pt}{1.3ex}}}

\begin{document}

\title{Theory of  Quantum Work in Metallic  Grains}

\author{Izabella Lovas}
\affiliation{Department of Physics and Institute for Advanced Study,
Technical University of Munich, 85748 Garching, Germany}
\affiliation{Munich Center for Quantum Science and Technology (MCQST), Schellingstr. 4, D-80799 M\" unchen}
\author{Andr\'as Grabarits}
\affiliation{BME-MTA Exotic Quantum Phases 'Lend\"ulet' Research Group, Institute of Physics, Budapest University of Technology and Economics, 
Budafoki \'ut 8., H-1111 Budapest, Hungary}
\author{M\' arton Kormos}
\affiliation{MTA-BME Quantum Dynamics and Correlations Research Group, 
Institute of Physics, Budapest University of Technology and Economics, 
Budafoki \'ut 8., H-1111 Budapest, Hungary}
\author{Gergely Zar\'and}
\affiliation{BME-MTA Exotic Quantum Phases 'Lend\"ulet' Research Group, Institute of Physics, Budapest University of Technology and Economics, 
Budafoki \'ut 8., H-1111 Budapest, Hungary}
\affiliation{MTA-BME Quantum Dynamics and Correlations Research Group, 
Institute of Physics, Budapest University of Technology and Economics, 
Budafoki \'ut 8., H-1111 Budapest, Hungary}


\begin{abstract}
We generalize Anderson's orthogonality determinant formula to describe  the statistics  
of work  performed on generic  disordered, non-interacting fermionic nanograins during quantum quenches.  The energy absorbed increases linearly with time, while its variance exhibits 
 a superdiffusive behavior due to Pauli's exclusion principle.    The probability of adiabatic evolution
decays as a stretched exponential.  In slowly driven systems, work statistics exhibits universal features, 
and can be  understood in terms of   fermion diffusion in energy space,
 generated by Landau-Zener transitions.  This diffusion is very well  captured  by a Markovian symmetrical exclusion process,  
 with the diffusion constant identified as the energy absorption rate.  The energy absorption rate shows an anomalous 
frequency dependence  at small energies, reflecting the symmetry class of the underlying Hamiltonian. 
Our predictions can be  experimentally verified  by calorimetric measurements performed on nanoscale circuits.
\end{abstract}

\maketitle

\section{Introduction} 
While energy transfer,  heat,  and work    are fundamental concepts in thermodynamics and statistical physics, 
it is far from trivial to extend their concepts to generic non-equilibrium quantum systems~\cite{workreview,Hanggi}, 
where energy becomes a fluctuating statistical quantity even in pure quantum states,
 and energy transfer can only be understood in terms of  a precise measurement protocol. 
Recent experimental developments allow, however, the investigation of these notions   in quantum systems ranging from individual molecules subject to mechanical forces~\cite{bioreview,molecule1,molecule2},  through nuclear spins in a magnetic field~\cite{batalhao}, 
to  mesoscopic  grains~\cite{pekola,pekola2}, and allow  even to extract the   \textit{full distribution} of the energy transferred.

\begin{figure}[b!]
\includegraphics[width=1\columnwidth]{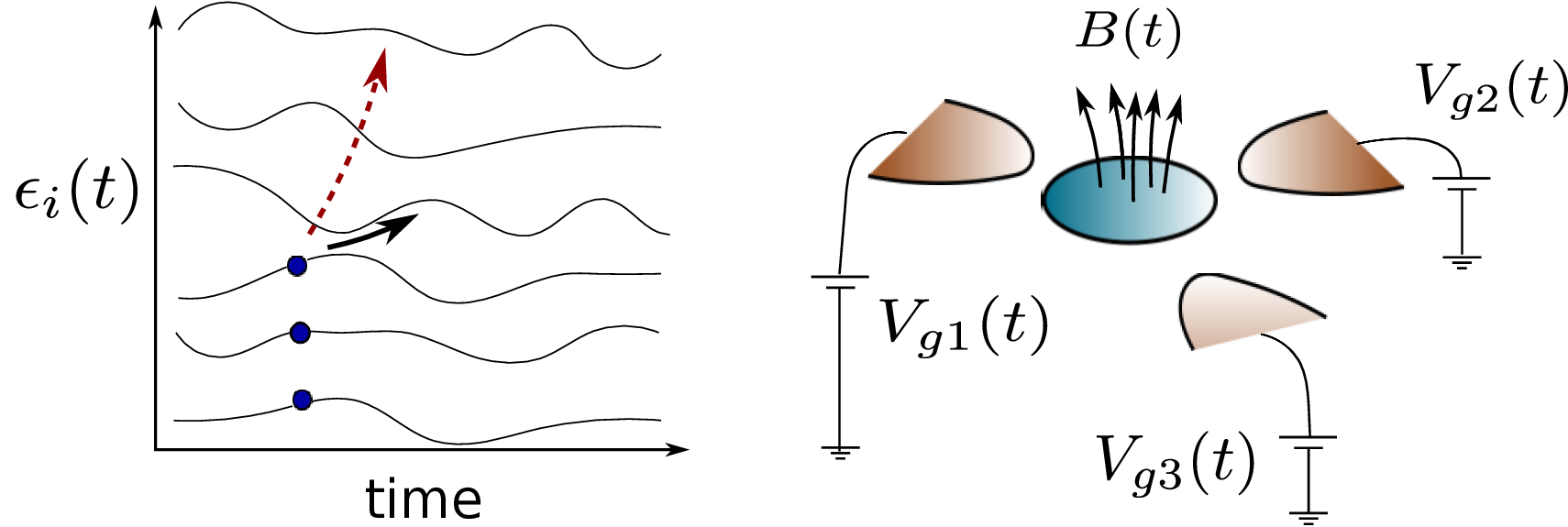}
\caption{
Quantum quenches for electrons in a generic disordered grain. 
(a) The non-interacting fermions occupy  levels of a random Hamiltonian. 
(b) External time dependent gate voltages and fields move the levels and induce  transitions between them.
For slow changes, these happen through  Landau-Zener transitions between neighboring levels (black arrow in panel (a)), 
while for faster changes electron-hole excitations between remote levels dominate (red dashed arrow in panel (a)).}
\label{fig:sketch}
\end{figure}

Along with this amazing experimental progress,  exact fluctuation theorems
have been derived and experimentally verified~\cite{workreview,jarz,crooks,molecule1,molecule2,batalhao,pekola,pekola2},   links between energy transfer (work), Lochschmidt echo, and quantum information scrambling has been established~\cite{otocGu,otoc1,otocexp,otocexp2}, and  the full distribution of work has been investigated in   many-body systems such as  Luttinger liquids~\cite{lutt,FD,FD1,FD2,liebliniger} or systems close to quantum criticality~\cite{gap1,gap2}.
So far, however, only very few studies investigate the effect of randomness~\cite{random1,random2,random3},  playing 
a crucial role in most nanosystems, and even these studies focus on  sudden quenches and do not address quantum statistics and/or interactions.

In this work,  we study the properties and universal aspects of   quantum work produced in course of time dependent quantum quenches
in  generic, non-interacting, but disordered fermionic many-body nanosystems.  These systems provide an ideal  platform to study 
quantum thermodynamics and the  interplay of quantum time evolution, disorder, and quantum statistics.

Generic noninteracting nanosystems 
obey random matrix statistics~\cite{matrixreview,RMreview}, and  can be  described in terms of 
a  Hamiltonian
\begin{equation}\label{eq:H}
\hat{H}(t)=\sum_{i,j=1}^N \hat{a}_i^\dagger\,\mathcal{H}_{ij}(t)\,\hat{a}_j\,,
\end{equation}
with  the $\hat{a}_i$ denoting fermionic annihilation operators, and  $\mathcal{H}(t)$  an $N\times N$ 
random matrix.  We assume that the system is in its $M$-particle ground state at time $t=0$, 
and then work is performed  by changing  external gate voltages or by applying time dependent magnetic 
fields~\cite{mesobook} (see Fig.~\ref{fig:sketch}). We describe this situation by performing a quench (ramp)  at a constant 
pace, $v\equiv \langle\text{Tr}( \text{d}_t {\cal H}^2)\rangle^{1/2}$,  related   to the frequency $\omega$ of  external parameters,
and investigate the distribution of the internal energy injected, 
\begin{equation}
P_t(W)\equiv \Big\langle\Big\langle \delta\big [W-(\hat H(t)-E_\text{GS}(t))\big]\Big\rangle\Big \rangle_{\rm RM},
\label{eq:pdf_def}
\end{equation}
with the two averages referring to quantum and random matrix averages, respectively.  
Though work and heat cannot be quite disentangled in course of 
the quench  process, throughout this paper  we follow the standard practice~\cite{workreview} 
and refer to this internal energy change  --
directly related to calorimetric measurements in driven nanosystems~\cite{calorimetry1,calorimetry2} -- 
as \emph{work}.

Following Ref.~\cite{wilkinson2},  we apply a  quench 
protocol, $\mathcal{H}(t)=\mathcal{H}_1\cos\lambda(t)+\mathcal{H}_2\sin\lambda(t)$, with $\dot{\lambda}$ set to constant, and  
   $\mathcal{H}_{1,2}$   independent $N\times N$  matrices, 
drawn from a Gaussian random matrix ensemble~\cite{matrixreview,footnote_J},
$\mathcal{P}\left(\mathcal{H}\right)\sim e^{-\beta\frac{ N}{4 }{\rm Tr}\mathcal{H}^2}$.
Here we focus on Gaussian orthogonal ($\beta=1$) and Gaussian unitary  ($\beta=2$) 
ensembles, corresponding to integer spin time reversal invariant systems, and systems 
with broken time reversal symmetry, respectively.

We first construct a determinant formula for the generating function of work in non-interacting 
fermionic systems. The determinant  formula presented can be considered as the dynamical analogue 
of Anderson's determinant formula for  the orthogonality catastrophe\cite{Anderson}.  However, while Anderson  restricts himself to the probability of staying in the 
ground state after a local sudden quench (placing a scatterer at the origin), the determinant formula used here 
describes  the response to any dynamical quench in a non-interacting fermionic many-body system, and describes all 
possible transitions and the corresponding overlaps.

We find that  energy is absorbed by the system via  hard core \emph{particle diffusion} in \emph{energy space}, 
and that the  statistical properties of work depend crucially  on the speed $v$ of the quench as well as on underlying symmetries. 
For slow, almost  adiabatic   changes, in particular,   $P_t(W)$ displays a universal structure, with surprisingly 
large, superdiffusive work fluctuations,   $\langle \delta W^2\rangle\sim \langle W\rangle^{3/2}$.   Energy absorption 
at small frequencies, $\omega \sim v$, reflects the symmetry (universality  class) of the Hamiltonian and
is predicted to scale as, $\langle W\rangle\sim \omega^{1+\beta/2}$,  with the random matrix parameter 
$\beta=1,2$ and $4$  corresponding to  orthogonal, unitary, and symplectic 
symmetries.

 The paper is organized as follows. We discuss the time dependent many-body wave function and present a determinant formula for the characteristic function of work in Sec. \ref{sec:psi}. We study the full distribution of work in Sec. \ref{sec:pdf}. In Sec. \ref{sec:average} we turn to the average work and show that it can be understood in terms of a diffusion in energy space. To shed more light on the dynamics, in Sec. \ref{sec:Markov} we show that the most important features of the work statistics can be captured by  a classical symmetric exclusion process, and in Sec. \ref{sec:MF} we also present a simple mean field description allowing the derivation of analytical results. We summarize our main findings in Sec. \ref{sec:discuss}. Technical details of the calculations, as well as additional numerical results for the Gaussian unitary ensemble are relegated to appendices.

\section{Many-body wave function}\label{sec:psi}
To compute the distribution \eqref{eq:pdf_def}, we  first need to determine the many-body wave function
 of our electron system after the quench, $|\Psi(t)\rangle$.  To do that, we exploit the fact  
 that our Hamiltonian is non-interacting, and  
 --  the initial state of the system being a Slater determinant --   the state $|\Psi(t)\rangle$ 
 can also be expressed as a Slater determinant at any time in terms of the time-evolved 
 single particle wave functions. Alternatively, $|\Psi(t)\rangle$ can be written as
\begin{equation}\label{eq:psi}
|\Psi(t)\rangle=\prod_{m=1}^{M} \hat{c}_{m,t}^\dagger |0\rangle
\end{equation}
with the operators $\hat{c}_{m,t}^\dagger$ creating  single particles in the states 
${\varphivec}^m(t)$;    $\hat{c}_{m,t}^\dagger|0\rangle\equiv| \varphivec^m(t)\rangle$. 
Here the vectors $\varphivec^m(t)$ satisfy the single particle 
Schr\" odinger equation 
\begin{equation}
\label{eq:sppsi}
i\,\partial_t\varphivec^m(t) ={\cal H}(t) \varphivec^m(t)
\end{equation}
with the boundary conditions $\varphi_i^m(0)= \delta^m_i$. 

To solve Eq.~\eqref{eq:sppsi},  we use the adiabatic approach. We 
introduce the  instantaneous eigenvectors  of ${\cal H}(t)$,  $\etavec_{\,t}^m $,
satisfying ${\cal H}(t)\, \etavec_{\,t}^m= \varepsilon_m(t) \etavec_{\,t}^m    $, and 
 corresponding fermionic creation operators, 
$|\etavec_{\,t}^m\rangle\equiv\hat{b}_{m,t}^\dagger|0\rangle$. 
We then expand the  $\varphivec^m(t)$'s   in the 
instantaneous basis as 
\begin{equation}
\varphivec^m(t) =\sum_k\alpha^m_k(t) \,\etavec^k_t\,,  
\end{equation}
and determine  the expansion coefficients by solving the corresponding equation of motion, 
\begin{align}
&i \,\dot{\alpha}_k(t)= \varepsilon_k(t)  \alpha_k(t) + \sum_{l} A^{kl}(t)\,\alpha_l(t)
\label{eq:schrodinger}
\end{align}
with $A^{kl} 
 = -i \; \etavec^k_t  \cdot \partial_t \etavec^l_t$
the Berry connection~\cite{footnoteA} (for details of the numerical calculation see Appendix \ref{sec:timeevol}). 
 
Since   $\hat c^\dagger_{m,t } = \sum_k \alpha^m_k(t) \,\hat b^\dagger_{k,t} $,  knowledge of 
the coefficients $\alpha^m_k(t)$  
 allows us to express the many-body state  $|\Psi(t)\rangle$  in   the instantaneous  basis.  
Observing furthermore that the  many-body Hamiltonian assumes a particularly simple form in the instantaneous basis, 
$\hat H(t) = \sum_m \varepsilon_m(t) \,\hat{b}_{m,t}^\dagger\hat{b}_{m,t}$, 
we can rewrite the expectation value  $\langle\Psi(t)|\, e^{iu\hat{H}(t)}\,|\Psi(t)\rangle$ as 
\begin{widetext}
\begin{align}\label{eq:guaux}
\langle\Psi(t)|\, e^{iu\hat{H}(t)}\,|\Psi(t)\rangle=&\sum_{\substack{\{k_1,k_2,...,k_{M}\},\\\{k_1^\prime,k_2^\prime,...,k_{M}^\prime\}=1}}^N e^{iu\sum_{m=1}^{M}\varepsilon_{k_m}(t)}\; \prod_{m=1}^{M} \alpha_{k_m}^m(t)\left[\alpha_{k_m^\prime}^m(t)\right]^*\langle 0|\hat{b}_{k_1^\prime,t}\hat{b}_{k_2^\prime,t}...\hat{b}_{k_{M}^\prime,t}\hat{b}_{k_{M},t}^\dagger...\hat{b}_{k_2,t}^\dagger\hat{b}_{k_1,t}^\dagger|0\rangle.
\end{align}
\end{widetext}
We can now evaluate the expectation values using Wick's theorem 
and eliminate the summation  over the labels $\{k_1^\prime,k_2^\prime,...,k_{M}^\prime\}$.
Transferring furthermore the permutation operators from the lower index of the $\alpha_k^m$'s to their upper 
index and reordering sums and products yields
\begin{align}
\langle e^{iu\hat{H}(t)}\rangle=&
\sum_P (-1)^P  \prod_{m=1}^{M} \left[g_t(u)\right]^{m,Pm}
 \end{align}
with $P$ running over all permutations of the $M$ occupied states, and the $M\times M$ matrix $g_t(u)$ incorporating all information on the overlap between initial and final single particle states,
\begin{equation}
 \label{eq:g}
\left[g_t(u)\right]^{m m^\prime} \equiv \sum_k [\alpha_k^m(t)]^*\,e^{i\,u\,\varepsilon_k(t)}\,\alpha_k^{m^\prime}(t).
\end{equation}
This leads  immediately to the  dererminant formula,
\begin{align}
G_t(u)&=
\big\langle  \big\langle\Psi(t)|\, e^{iu\left(\hat{H}(t)-E_\text{GS}(t)\right)}\,|\Psi(t)\big\rangle\big\rangle_\text{RM}\nonumber\\
&=\big\langle e^{-i\,u\,\sum_{m=1}^{M}\varepsilon_m(t)}\,{\rm det}\, g_t(u)\big\rangle_\text{RM}\,.
\label{eq:Gu}
\end{align}
Eq.~\eqref{eq:Gu} is one of the important results of this work: it establishes a connection between the work statistics of the many-body system and 
the time evolution of individual single particle states~\cite{Poster,FeiQuan}. 
%
In the following, we shall use this formula to study the properties of 
quantum quenches in generic fermionic nano-systems described by random matrix theory.

\begin{figure*}[t!]
\includegraphics[width=1\textwidth,clip]{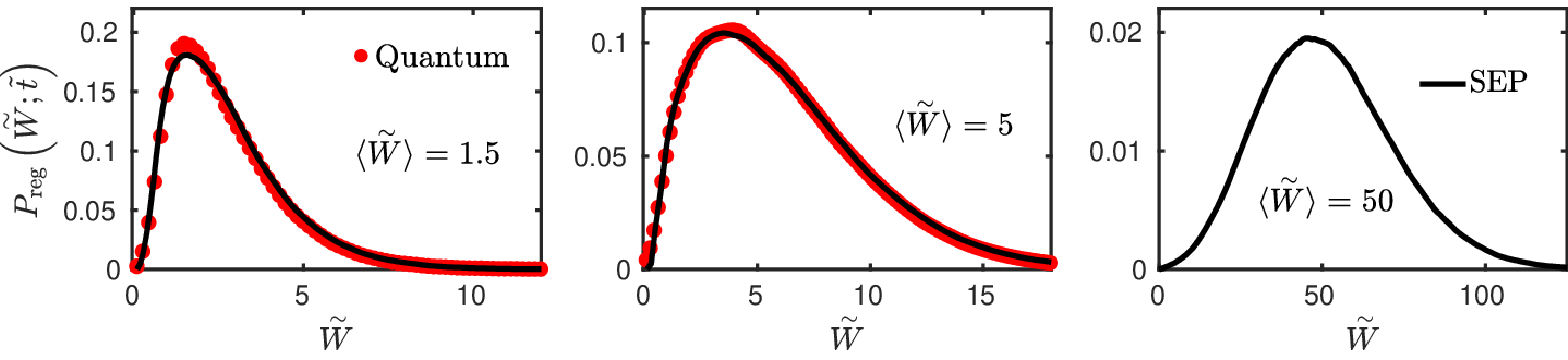}
\caption{
Time evolution of work statistics for the Gaussian orthogonal ensemble  in the limit, $\tilde v\lesssim 1$. Circles: quantum results for the regular part, $P_{{\rm reg}}(\Wtilde; \tilde{t})$, for dimensionless velocity $\tilde{v} = 0.8$ and  $N = 20$ levels. In the three panels, we used different dimensionless quench times  $\tilde{t}=4.88,\,6.25$, and 162.5 (from (a) to (c)), corresponding to dimensionless average work $\langle \Wtilde \rangle=1.5,\,5$, and 50, respectively~\cite{footnoteB}.  
Solid lines:  SEP simulations.  The distribution becomes more and more Gaussian-like
as $\langle \Wtilde \rangle$ increases. 
}\label{fig:P(W)}
\end{figure*}

\section{Quantum statistics of work}\label{sec:pdf}
\begin{figure}[b!]
\includegraphics[width=0.9\columnwidth,clip]{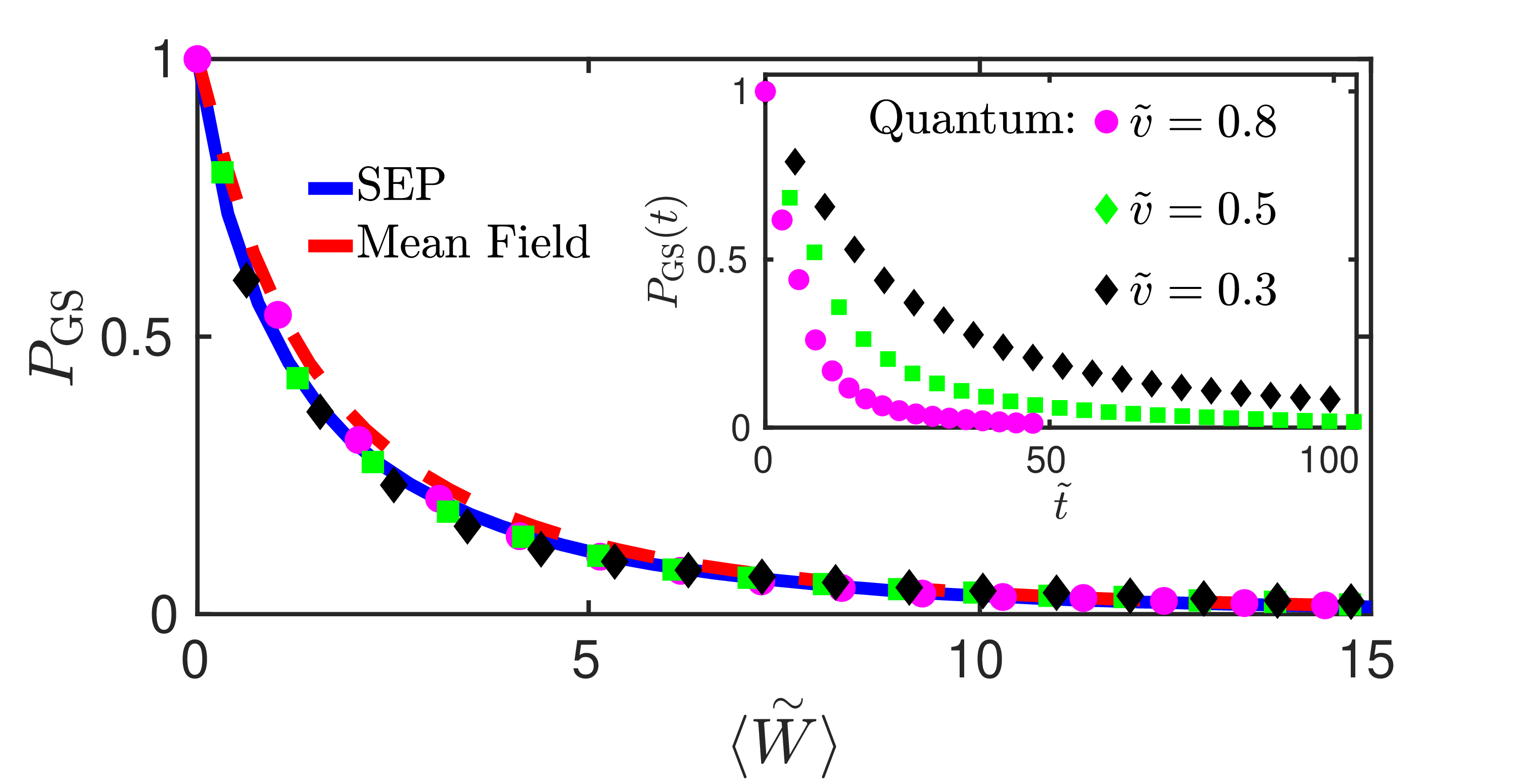}
\caption{Probability of adiabaticity.
Inset:  Time evolution of the adiabatic process probability, $P_{{\rm GS}}$, for the orthogonal ensemble ($\beta = 1$)
for three different velocities, $\tilde v\lesssim 1$. 
Main panel: Symbols: same data as in inset, but plotted as functions of  $\langle \Wtilde \rangle$, 
 collapsing all to a single curve. Solid line: classical SEP result. Dashed line: {Eq. \eqref{eq:PGS_MF} obtained by} mean field theory. 
}\label{fig:P_GS}
\end{figure}

To determine the full distribution $P_t(W)$, we utilized   Eqs.~\eqref{eq:schrodinger} and \eqref{eq:Gu}.  
We generated random matrices  ${\cal H}_1$ and ${\cal H}_2$,  determined  the expansion coefficients 
$\alpha^m_k(t)$ and the determinant ${\rm det}\, g_t(u) $ numerically, averaged over the random matrix 
ensemble, and finally determined $P_t(W)$ by taking the Fourier transform of Eq.~\eqref{eq:Gu}.
In our numerical simulations,  we focus on the half-filled case, $M=N/2$, though the results 
are independent of this assumption and carry over to any filling  $M/N$.
Our results are summarized in  Figs.~\ref{fig:P(W)}  and  \ref{fig:P_GS}. 

The function $P_t(W)$ can be disentangled into an adiabatic (ground state) and 
a regular part as    
\begin{align}\label{eq:preg}
P_t(W)=P_\text{GS}(t)\, \delta(W)+P_{\rm reg}(W;t).
\end{align}
These functions depend not only on time, but also on the  velocity  
$v$,
by which we drag the Hamiltonian through the random matrix manifold,  on the position 
of the Fermi level, and on the symmetry class, too. 
However, once expressed in terms of appropriate dimensionless 
quantities, $\Wtilde \equiv W/\Deltabar$, $\tilde t\equiv  t\,\,\Deltabar$, and 
$\tilde{v}\equiv {v}/ \Deltabar$, measured in units of the average single particle level spacing
$\Deltabar$ at the Fermi energy, they become \emph{universal} functions 
in the limit $N\to \infty$, $P_{\rm reg}\to P_{\rm reg}(\Wtilde;\tilde t)$  and $P_\text{GS}\to P_\text{GS}(\tilde t)$, displayed in 
Figs.~\ref{fig:P(W)}  and  \ref{fig:P_GS} for the   orthogonal ensemble. They depend implicitly on $\tilde v$ but, supposedly, they are  independent of all microscopic details, and depend just on the symmetry of the underlying Hamiltonian (see also Appendix \ref{sec:universality} for data for the Gaussian unitary ensemble).  
Remarkably, these functions simplify even further in the slow quench limit, $\tilde v\lesssim 1$,
and become functions only of the average work,  $\langle \Wtilde \rangle$, rather than time and velocity. 
This is demonstrated for the probability of adiabatic transitions in Fig.~\ref{fig:P_GS}
within the orthogonal ensemble.

The  functions $P_{\rm reg}(\Wtilde;\tilde t)$ show a rather slow evolution towards a Gaussian 
distribution, as we increase $\langle \Wtilde\rangle.$  As closer 
numerical investigation reveals, the work fluctuations follow a superdiffusive scaling, ${\delta\!\Wtilde{}\!\!^2\sim \tilde t^{\,3/2} \sim \,\Wtilde{}\!\!^{\,3/2}}$. 
The behavior of $P_\text{GS}(\tilde t)$ is also somewhat unexpected: the probability of 
an adiabatic transition is found to decay as a stretched exponential,  $P_\text{GS}(\tilde t)\sim{\tilde t}^{\,1/4}\, e^{-C\sqrt{\tilde t}}$.  As we discuss and demonstrate below, both are  particular features of  a symmetrical 
 exclusion process which governs  the dynamics in energy space.

\section{Energy space diffusion and average work}\label{sec:average}
\begin{figure}[b!]
\includegraphics[width=0.9\columnwidth]{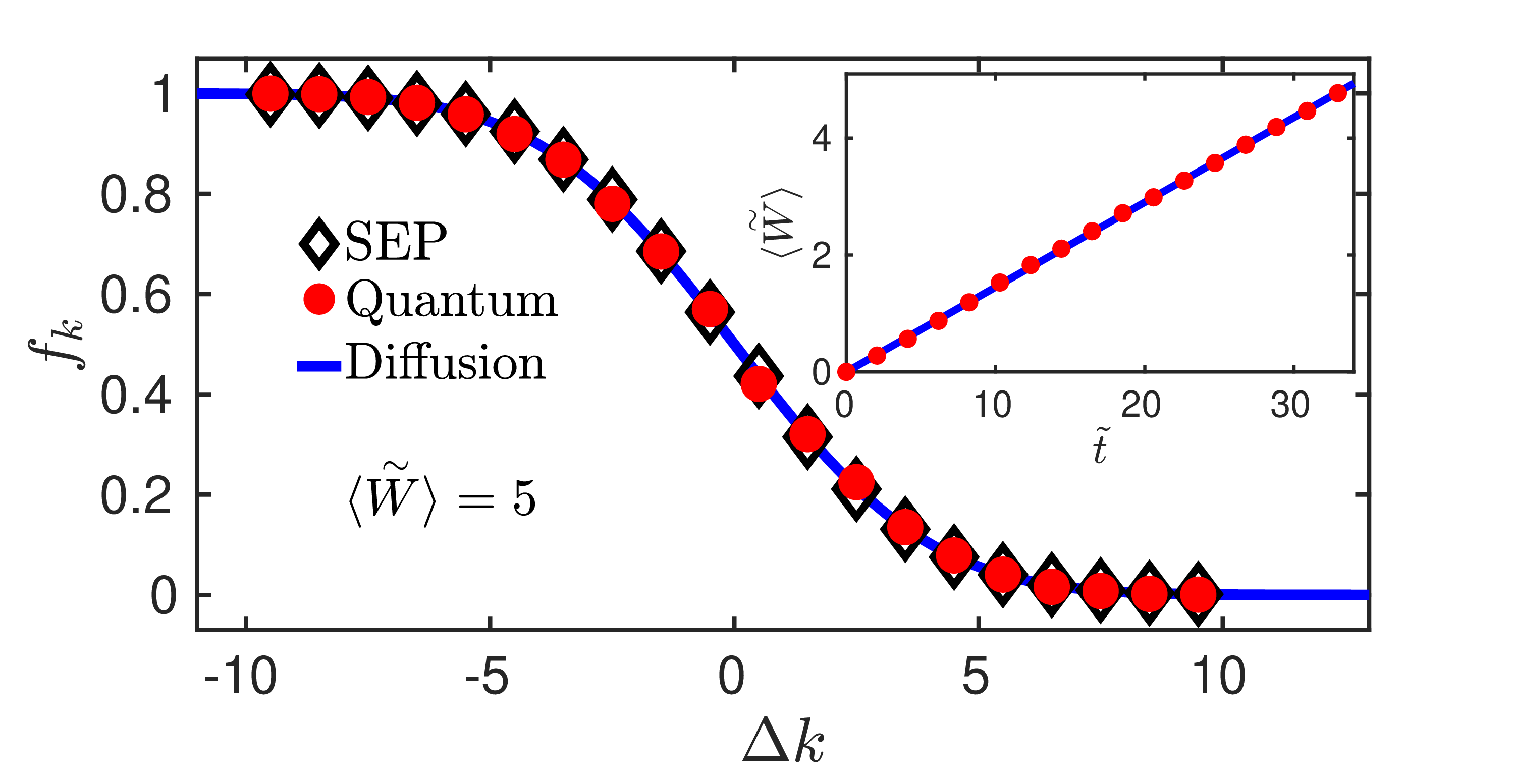}
\caption{
\label{fig:diffusion}
Diffusive broadening of the occupation profile  $f_k(t)$ for the orthogonal ensemble, plotted as a function of $\Delta k = k - M$, 
for $\langle \Wtilde \rangle = 5$.
 Circles: quantum results  for a dimensionless velocity $\tilde{v} = 0.5$ and $N = 20$. Diamonds: SEP   simulations.  Solid line:  diffusive error function fit. Inset: time evolution of  dimensionless average work, $\langle \Wtilde\rangle$ (circles), as a function of the dimensionless time, $\tilde{t}$.  Solid line: SEP simulation yielding ${\langle \Wtilde\rangle\approx \widetilde D \,\tilde t,}$  with a dimensionless diffusion constant $\widetilde{D} = 0.145.$ }
\end{figure}

Fig.~\ref{fig:diffusion} shows the average of the occupation of each level 
$ f_{ k}(t)\equiv \langle\langle \hat n_{k,t}\rangle\rangle_\text{RM}$.\cite{footnote_f_k}
The occupation profile exhibits a clear diffusive character, and is very precisely  described 
by a diffusively broadened Fermi-sea, 
\begin{equation}\label{eq:fk}
 f_{ k}(t)\approx \left[1-
{\rm erf}\left(    \Delta k/( 4 \widetilde{D}\tilde{t})^{1/2}\right) \right] /2,
\end{equation}
where
 $\Delta k \equiv k-M$ is  the distance of level $k$  from the Fermi energy, and
$\widetilde D$ denotes a dimensionless  diffusion constant in energy space. 
The diffusive $\sim \sqrt{t}$  broadening  of the Fermi surface immediately
implies a \emph{linear} internal energy absorption, $\sim t$, 
\begin{equation}
\label{eq:diffW}
\langle W(t)\rangle\approx \Deltabar \,\,\widetilde{D}(\tilde{v})\;\tilde{t}\;, 
\end{equation}
as clearly demonstrated in the inset of Fig.~\ref{fig:diffusion}. 
The diffusion constant $\widetilde{D}$ can thus be interpreted as an overall dimensionless 
energy absorption rate. 

The  constant $\widetilde D$ depends  on   the universality class of the system as well as on the 
 velocity, $\tilde v$.  We can distinguish two distinct regimes: 
in the ``fast limit", $\tilde v\gg 1$, electron-hole 
transitions between remote levels dominate energy absorption, 
and we find  $\widetilde D(\tilde v)\sim \tilde v^2\sim \omega^2$, as expected in metals. 
However, in the ``slow limit",  $\tilde v\lesssim 1$, nearest neighbor transitions mediated by Landau-Zener transitions
dominate. The statistics of these latter have been studied thoroughly~\cite{wilkinson1,wilkinson2,wilkinsondiff1}, 
and it has been shown that  the parameters of Landau-Zener transitions display 
universal distributions (see also Appendix \ref{sec:crossing} for details). 
A simple calculation making use of this universal statistics then yields 
 an energy absorption    $\widetilde D( \tilde{v})\sim  \tilde v^{1+\beta/2} \sim |\omega|^{1+\beta/2}$
   for  $\tilde{v}\lesssim 1$, as indeed confirmed by   our  detailed 
numerical simulations (see Appendix \ref{sec:crossing}).

\section{Markovian simulation and  symmetrical  exclusion process}\label{sec:Markov}
In the slow limit, $\tilde v\lesssim 1$, dominated  by Landau-Zener transitions, 
 most features of the work statistics can be understood in terms of a simple, 
classical  model, the symmetrical exclusion process (SEP) in energy space. 
In this approach, we consider the occupation of each level as a classical 
statistical variable, taking values $n_{k,t}=1$ and 0,  and think of 
Landau-Zener transitions  as random, Markovian events, transferring particles between 
neighboring levels with some probability $p_\text{LZ}$. 
The probabilities $P_{\{n_k\}}$ can then be obtained by performing Monte Carlo averaging. First we initialize the fermions on the $M$ lowest levels, then we apply a Markov process in the space of level occupations. During this process, the fermion configuration can only change by nearest neighbor transitions and each such transition happens with rate $\tilde{D}(\tilde{v})$. Note that this model uses a single parameter extracted from the full quantum simulation, the diffusion constant $\tilde{D}(\tilde{v})$~\cite{footnoteC}.  We then  determine the work statistics as 
\begin{equation}
\label{eq:pdf}
P_t(W)=\Big\langle\sum_{\{n_k\}} P_{\{n_k\}}\, \delta\left[W-\big(E_{\{n_k\}}(t)-E_\text{GS}(t)\big)\right]\Big\rangle_{\rm RM}
\end{equation}
with $E_{\{n_k\}}(t) = \sum_k \varepsilon_k(t) n_k(t) $, and the random matrix average performed only on the 
final eigenenergies, $\varepsilon_k(t)$.

As shown in Figs.~\ref{fig:P(W)}, \ref{fig:P_GS}, and \ref{fig:diffusion}, 
apart from the very short time behavior where quantum mechanics rules, the simple SEP approach 
 reproduces 
the results of our fully quantum mechanical computations with amazing accuracy for slow quenches.
This result now allows us to use  SEP computations to obtain 
predictions for the work distribution function in the regime of large injected work, 
$\langle \Wtilde\rangle \gtrsim 20$, inaccessible to our quantum mechanical simulations. 
These results are shown in the rightmost panel of Fig.~\ref{fig:P(W)}.

\section{Mean field description}\label{sec:MF} 
The SEP model thus provides an accurate description of energy absorption for the most interesting, 
universal  regime, $\tilde v\lesssim 1$, but provides limited analytical understanding. 
We can construct, however, a simple mean field theory of work distribution, by assuming that 
the occupations  $n_k$ are classical binary variables with expectation values $f_{k}(t)=\langle n_k\rangle$, 
and that they  are independent -- apart from overall particle number conservation. 
This simple mean field theory incorporates three important ingredients: 
Pauli principle, particle number conservation, and the diffusive character of Fermi surface broadening.  Below we demonstrate that this approach can be used to obtain analytical assymptotic estimates for the probability of adiabaticity and for the variance of work.

\subsection{Probability of adiabaticity}

Within the mean field approach, we consider each  occupation number $n_k$
as a binary probability variable, having values  $ n_k=0$ and $1$.  In the simplest classical approximation neglecting all correlations between different levels, at time $t$ we can assign  the probabilities $f_k(t)$ and $1 – f_k(t)$ to $n_k = 1$ and $n_k = 0$, respectively, 
\begin{equation}
\label{eq:bernoulli}
p_{k,t}(n_k)=n_k\, f_{ k}(t) +(1-n_k)\big(1- f_{ k}(t)\big)\;.
\end{equation}
Here we set the expectation value of $n_k$, $f_k(t)$, in accordance with the diffusively broadened Fermi sea, Eq. \eqref{eq:fk}. To incorporate correlations at the lowest order, we supplement Eq. \eqref{eq:bernoulli} with a global constraint expressing particle number conservation, $\sum_k n_k = M$. For simplicity, here we focus on the half-filled case $M\equiv N/2$, though 
calculations can easily be extended to any value of $M$.
We enforce the constraint by inserting a Kronecker $\delta$ function into the joint 
probability distribution, 
\begin{align}\label{eq:PMF}
P\left(\{n_k\}\right)&=\dfrac{1}{\mathcal{N}_t}\prod_{k=1}^N p_{k,t}(n_k)\;
\delta_{N/2=\sum_{k} n_k}\nonumber\\
&=\dfrac{1}{\mathcal{N}_t}\int_{-\pi}^{\pi}\dfrac{{\rm d}\lambda}{2\pi}e^{i\lambda\sum_{k=1}^N\left(n_k -1/2 \right)}\prod_{k=1}^N p_{k,t}(n_k)\;.
\nonumber
\end{align}
Here ${\mathcal{N}_t}$ is a time dependent normalization factor, 
which can be estimated  by first carrying out the summation over $\{n_k\}$, and then applying the saddle point approximation as
\begin{align}
\mathcal{N}_t &\approx \int_{-\pi}^{\pi}\dfrac{{\rm d}\lambda}{2\pi} \prod_{j>0}\left[\cos^2(\lambda/2)+\sin^2(\lambda/2) \mathrm{erf}^2\left(\frac{j}{\sqrt{4\widetilde D \tilde t}}\right)\right]
\nonumber 
\\
&\approx
\left(8\pi\widetilde D\tilde t\right)^{-1/4}\;.
\nonumber\end{align}
The probability of staying in the ground state then reads
\begin{align}
P_\text{GS}&(t) = \dfrac{1}{\mathcal{N}_t}\prod_{k=1}^{N/2}  f_{k}(t) \prod_{k=N/2+1}^{N} \big(1- f_{k}(t)\big)
\nonumber
 \\
\approx & 
\dfrac{1}{\mathcal{N}_t} e^{2\sqrt{4\widetilde D\tilde t}\int_0^\infty\! {\rm d}x \log[(1+\mathrm{erf}(x))/2]} \;.
\nonumber
\end{align}
Here, again, an integral approximation has been made by assuming $\widetilde D\tilde t\approx\; \Wtilde\,  \gg 1$ and the saddle point approximation has been carried out, leading to the asymptotic estimate, 
 \begin{equation}
 \label{eq:PGS_MF}
 P_\text{GS}(t)=(8\pi\widetilde D\tilde t)^{1/4}\,e^{-C\sqrt{\widetilde D\tilde t}}
 \;, 
\end{equation}
yielding an excellent fit for $\langle \Wtilde \rangle>1$, as 
shown in Fig.~\ref{fig:P_GS}.

\subsection{Variance of work}

For $\Wtilde\,\gg 1$, we can estimate the variance of the work by 
neglecting the fluctuations  of the individual energy levels~\cite{vonDelft_KondoBox_paper}, 
 $\varepsilon_k(t) \to  \Deltabar \, k$, 
 while keeping track of the fluctuations of the occupation numbers.
For a given realization of ${\cal H}(t)$, this yields the approximate expression
\begin{equation}
    \delta\!\Wtilde{}\!\!^2 (t) \approx 
    \Big\langle \big( \sum_{k = 1}^{N} \Delta k\,  \hat n_{k,t}\big)^2\Big\rangle 
    - \Big\langle \sum_{k = 1}^{N} \Delta k\,  \hat n_{k,t} \Big\rangle^2,
    \nonumber
\end{equation}
the average signs denoting here
just quantum averages. By separating the `diagonal' contributions, 
the average $\langle  \delta\!\Wtilde{}\!\!^2\rangle_{\rm RM}$  
can be rewritten as 
\begin{align}
    \langle  \delta\!\Wtilde{}\!\!^2& (t)\rangle_{\rm RM}  \approx 
   \sum_{k }\Delta k^2\,  \big\langle \big\langle \delta  \hat n_{k,t}^2\big\rangle  \big\rangle_{\rm RM}
   \nonumber
   \\
   & +  \sum_{k \neq k'}  \Delta k\,\Delta k'
     \big\langle\big\langle \delta  \hat n_{k,t}\delta  \hat n_{k',t}
     \big\rangle \big\rangle_{\rm RM} 
\label{eq:variance}
\end{align}
with $\delta  \hat n_{k,t} \equiv   \hat n_{k,t} -\langle   \hat n_{k,t}\rangle$ denoting the deviation 
of the occupation number. 
Since the  $\hat n_{k,t}$ behave as binary variables, averages in the first term  
can be expressed as 
$ \langle \langle\delta  \hat n_{k,t}^2\rangle\rangle_{\rm RM} = f_k(t) \big(1-  f_k(t) \big)$, 
with  $f_k\left(t\right) \approx \big(1 - {\rm erf}(\Delta k/\sqrt{4\widetilde{D}\tilde{t}}\,)\big)/2$. 
The correlators appearing in this equation  can be expressed in terms of the amplitudes $\alpha^m(t)$
 as  $\langle \delta \hat n_{k,t}\delta \hat n_{k',t}\rangle =
-  \big|\sum_{m=1}^{N/2} {\alpha_k^m (t)}^* \alpha_{k'}^m(t)\big|^2$.
Notice that this correction is negative, indicating that the level occupations are
\emph{anticorrelated}, as dictated by particle number conservation.

By neglecting for a moment these correlations and replacing sums by integrals, we obtain the estimate
\begin{equation}
 \langle  \delta\!\Wtilde{}\!\!^2 (t)\rangle_{\rm RM}  \approx 
\int_{-\infty}^{\infty} {\rm d}x\, x^2\, \frac{1 - {\rm erf}^2(x / \sqrt{4\widetilde{D}\tilde{t}})}{4}  \sim \tilde{t}^{3/2}\,,
\nonumber
\end{equation}
reproducing the superdiffusive scaling $ \langle  \delta\!\!\Wtilde{}\!\!^2 \rangle  \sim \langle \Wtilde\rangle^{3/2}$.
We note that while neglecting the correlations  explains   the observed behavior of the work's variance,
 it does not reproduce   the correct prefactor. Indeed, a more careful mean field calculation 
 along the lines of the  previous subsection shows\cite{MF} that  correlations (the conservation of fermion number)  
 cannot be entirely  neglected,    and they also give a similar but smaller $\sim \tilde t^{3/2}$ contribution 
 to the variance -- without changing the overall scaling.

\section{Conclusions}\label{sec:discuss}
We have derived and used a determinant formula  to compute 
the distribution of work, $P_t(W)$, in course of a quantum quench in generic, disordered, fermionic nano-systems, and have 
shown that it  displays a large degree of universality, especially in the slow quench limit.
Even if a complete experimental characterization   of $P_t(W)$ may  still be challenging, 
 many of our curious findings such as the diffusive broadening of the Fermi energy,  
 the $\sim t^{3/2}$ scaling of the variance of the energy absorbed, or the low frequency $\sim \omega^{1+\beta/2}$ 
 absorption rate, could be readily verified experimentally.

Our  results  also  demonstrate that  quantum work statistics in these systems   is, after all, to a large extent classical.
In the most interesting slow quench limit, we demonstrate a close connection to the symmetrical exclusion process (SEP), 
a \emph{classical  diffusion} of hard core  particles in energy space. 
Quantum mechanics enters here  through level collisions,  giving rise to  
Landau-Zener transitions,   the exclusion process,  mirroring the Pauli principle and, finally, 
 level statistics,  reflecting the symmetry of the underlying quantum mechanical system.

Though diffusion in energy space and its relation to thermalization  has been studied by several authors in 
different contexts\cite{wilkinsondiff1,JarzynskiPRE1993,CohenPRL2013}, earlier works have not focused on the 
impact of quantum statistics on energy diffusion. In particular, here we find that energy absorption 
in this many-body system corresponds to diffusion of \emph{strongly interacting} particles. 
This very strong interaction, captured at the classical level by the exclusion process, 
implies that while the total energy absorbed  shows a clear drift in time, $\langle W\rangle \sim t$, 
its spread is \emph{non-diffusive} but rather increases superdiffusively as $\langle\delta W^2\rangle \sim t^{3/2}$ during 
a quench.

  Our investigations, focusing on $T=0$ temperature non-interacting systems,
 represent clearly only the first step. Inclusion  of interactions,  
generalizations to the symplectic universality class as well as to finite temperatures and open systems 
are all  exciting open questions for future research. 

\begin{acknowledgments} We thank Jukka Pekkola and Bal\'azs D\'ora for insightful  discussions.  
This work has been supported by the National Research, Development and Innovation Office (NKFIH) through 
 Grant No. SNN 118028,  through  the Hungarian Quantum Technology National Excellence Program, 
 project no. 2017-1.2.1-NKP-2017- 00001,  by the BME-Nanotechnology FIKP grant  (BME FIKP-NAT),
and by the European Research Council (ERC) under the European Unions Horizon 2020 research and innovation program (grant agreement No.  771537). M. K. acknowledges support from a ``Pr\'emium'' and a ``Bolyai J\'anos'' grant of the HAS, and was supported by the \'UNKP-19-4 New National Excellence Program of the Ministry for Innovation and Technology.
\end{acknowledgments}

\appendix

\section{Statistics of avoided level crossing}\label{sec:crossing}

 \begin{figure}[b!]
\includegraphics[width=0.95\columnwidth,clip]{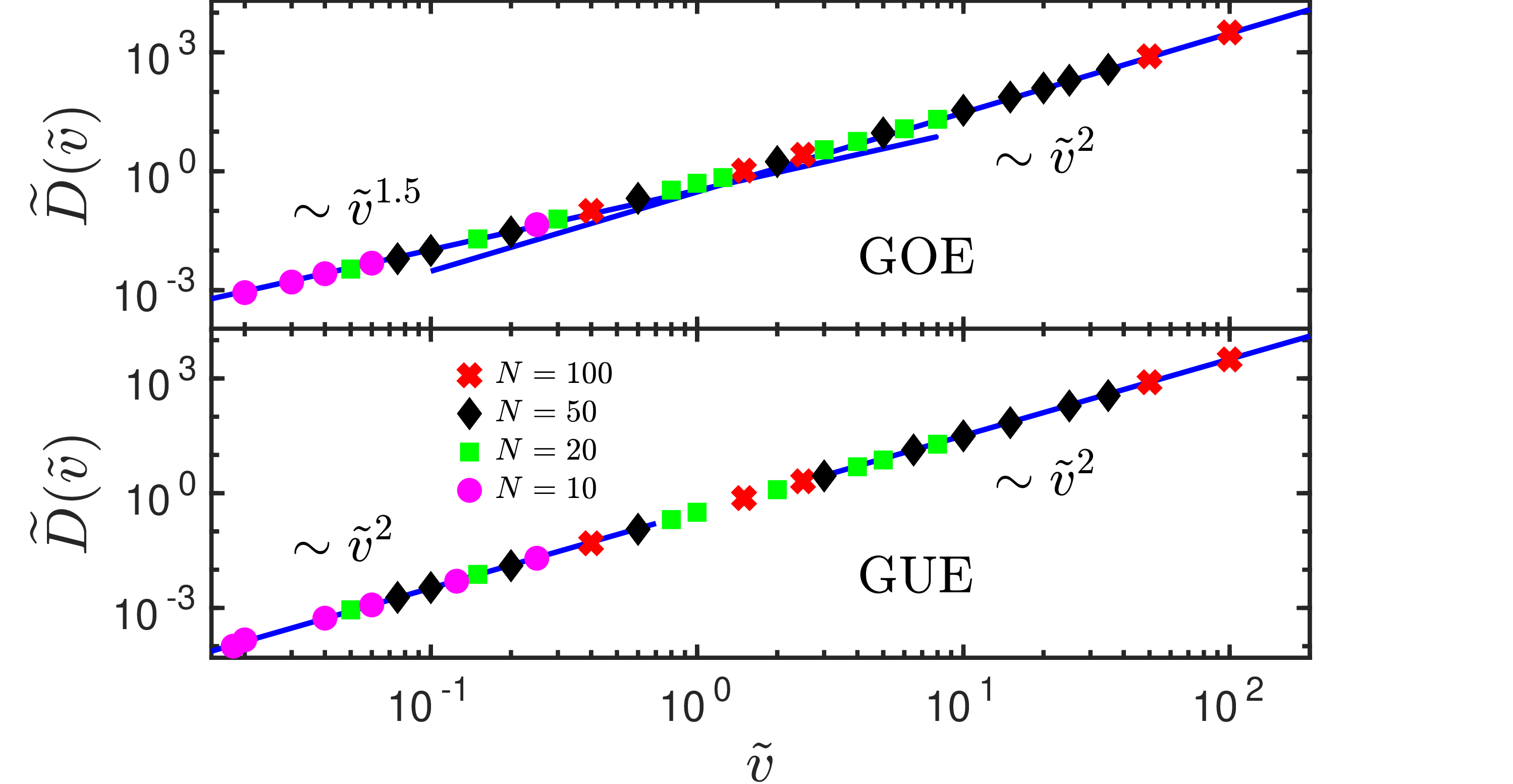}
\caption{
Universality of diffusion constants. Dimensionless diffusion constants plotted as the function of the dimensionless velocity of the quench, $\tilde{v}$, for (a) GOE  and (b) GUE, using different matrix sizes $N$, with logarithmic scale on both axes. The data trace out a universal curve, scaling as $\widetilde{D} \sim \tilde{v}^{1+\beta / 2 }$ in the slow quench limit, $\tilde{v} \lesssim1$. For fast quenches, $\tilde{v} \gg 1$, the statistical aspects of level repulsion lose their role and no longer govern the many-body dynamics, leading to $\widetilde{D} \sim \tilde{v}^2$ for both ensembles.
}
\label{fig:D_v}
\end{figure}

At small velocities, $\tilde v \lesssim1$, Landau-Zener transitions dominate the energy 
absorption process.  The statistical properties of these transitions are universal~\cite{wilkinson3,wilkinsondiff2}
and depend on the  particular universality class considered. In practice, we consider motions 
 within the random matrix manyfold along a `circle', 
\begin{equation}\label{eq:hl}
\mathcal{H}(\lambda)=\mathcal{H}_1\cos\lambda+\mathcal{H}_2\sin\lambda,  
\end{equation}
and investigate the separation of  neighboring levels, 
 $\Delta(\lambda)\equiv \varepsilon_{k+1}(\lambda)-\varepsilon_k(\lambda)$. 
This function displays well-defined minima, $\Delta_{\rm min}$, at positions $\lambda_0$, close to which  the level separation 
can be well-described in terms of an effective two-level  Hamiltonian as
\begin{equation}
\Delta(\lambda)\approx\sqrt{\Delta_{\rm min}^2+\gamma^2(\lambda-\lambda_0)^2}.
\label{eq:LZfit}
\end{equation}

\begin{figure*}[bth]
\includegraphics[width=0.65\textwidth]{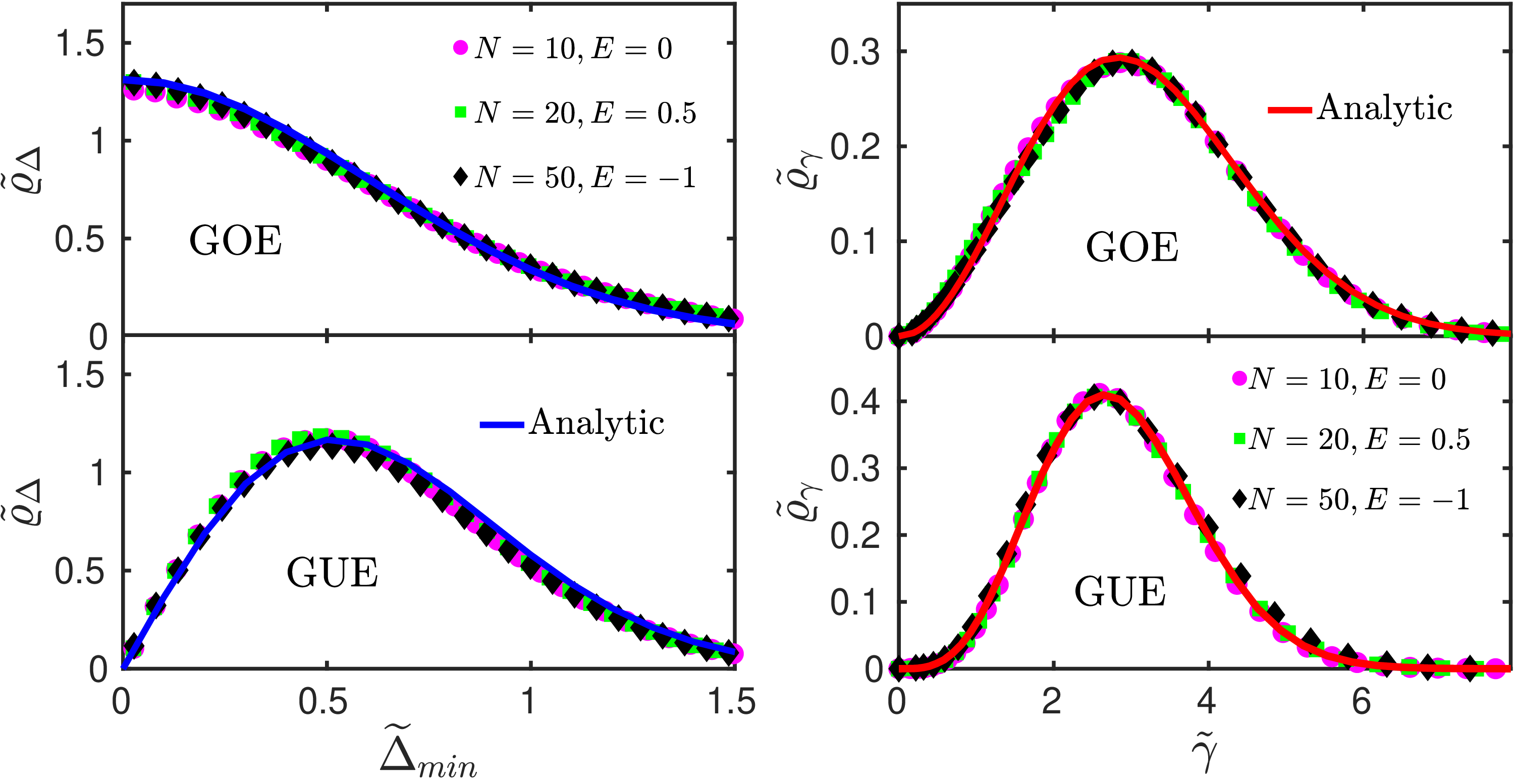}
\caption{
\label{fig:LZ_distributions}
Universality of distribution  of avoided level crossing parameters for the (a)-(b) Gaussian orthogonal (GOE) and (c)-(d) Gaussian unitary ensembles  (GUE).   (a),(c) The dimensionless Landau-Zener gap ${\widetilde \Delta}_\text{min}$, and (b),(d) slopes $\tilde{\gamma}$ display universal distributions, 
 depending only on the symmetry class of $\cal H$,   independent  of its  size, $N$, and the energy, $E$.  Notice that 
 $\tilde \varrho_{\Delta}({\widetilde \Delta}_\text{min})\sim {\widetilde \Delta}_\text{min}^{\beta-1}$, implying
 $\tilde\varrho_{\Delta}( 0) = \text{finite}$ for GOE,  $\beta=1$, and a breakdown of the adiabatic limit. 
  Continuous lines denote  analytical results of Refs.~\cite{wilkinson3,wilkinsondiff2}. 
  }
\end{figure*}

The number of level crossings occurring between two selected neighboring levels 
 is directly proportional to the distance covered within the random matrix manifold, 
 $N_{\rm cross}\sim s_\text{tot}\sim \sqrt{N} \,{\rm d}\lambda$. Therefore, we 
 can define the probability distribution of $\Delta_{\rm min}$ as 
\begin{equation}\label{eq:crossnro}
{\rm d}N_{\rm cross}=\rho(\Delta_{\rm min})\,{\rm d} s \,{\rm d}\Delta_{\rm min}\;.
\end{equation}
Here the density function $\rho(\Delta_{\rm min})$ depends implicitly on the size of the 
random matrices, $N$, as well as on the universality class considered~\cite{footnoterho}.
The dependence on $N$, however, appears just in a trivial way in the large $N$ limit, through the level spacing
$\Deltabar$ at the energy of the Landau-Zener transitions.  This dependence can be scaled out,
 yielding a universal distribution function,  
\begin{equation}\label{eq:rhodimless}
\tilde \varrho(\widetilde \Delta_{\rm min})\equiv {\Deltabar}\,\, {\rho}\left({\Delta}_{\rm min}\right),
\end{equation}
with $\widetilde \Delta_{\rm min} =  \Delta_{\rm min}/\Deltabar$ denoting the dimensionless Landau-Zener gap.

Similarly, the distribution of the slopes $\gamma$ is also universal, provided we measure it in its 
natural unit, 
$$
\tilde{\gamma}=\gamma \sqrt{N}.
$$
The numerically computed statistics of the distributions $\tilde\varrho(\widetilde \Delta_{\rm min})$ 
and $\tilde\varrho(\tilde \gamma)$ are displayed for the Gaussian orthogonal ensemble (GOE, $\beta=1$) and 
Gaussian unitary ensemble (GUE, $\beta = 2$) in Fig.~\ref{fig:LZ_distributions}. The numerically extracted distributions 
fit perfectly the analytical predictions of Ref.~\onlinecite{wilkinson3}. In particular, the 
 distribution of the slopes has a peaked structure, i.e., there exist  a typical value of $\tilde \gamma$, 
 characterizing the transitions.

 The  distribution $\tilde \varrho(\widetilde \Delta_{\rm min})$ is, on the other hand, more peculiar. 
In particular, $\tilde \varrho(\widetilde \Delta_{\rm min})\sim \Delta_{\rm min}^{\beta-1}$, and scales as
\begin{equation}\label{eq:mingap}
\tilde{\varrho}(\widetilde{\Delta}_{\rm min}\ll 1)\approx
\begin{cases}
\pi^{3/2}/\sqrt{18} & \text{for $\beta=1$, }
\\
{2\, \pi^{3/2}} \,\,\widetilde{\Delta}_{\rm min} /3
& \text{for $\beta=2$, }
\end{cases}
\nonumber
\end{equation}
in the orthogonal and unitary universality classes, investigated here in detail~\cite{footnote_pi}.  This curious behavior implies  
that  avoided level crossings with very small gap are abundant in the orthogonal  class, 
which leads to a breakdown of adiabatic perturbation theory for $\beta=1$.

The probability $p_{\rm LZ}$ of a Landau-Zener transition between two neighboring levels 
depends on the velocity at which the transition is approached, and is known analytically~\cite{LZ1,LZ2}, 
\begin{equation}
\label{eq:LZprob}
p_{\rm LZ}=\exp\left(-\dfrac{\pi^2}{2}\dfrac{\widetilde\Delta_{\rm min}^2}{\tilde{v}\,\tilde{\gamma}}\right)\,.
\end{equation}
Given the exponential sensitivity to $\Delta_{\rm min}$, small Landau-Zener gap transitions dominate 
in the limit of small velocities, $\tilde v\ll1$. This leads to the estimate 
\begin{align}
\label{eq:DLZ}
\widetilde{D}&\sim\big \langle {\rm d}_t N_{\rm cross} \,\, p_{\rm LZ}\big\rangle=\tilde{v}\int {\rm d}\widetilde{\Delta}_{\rm min} \,\tilde\varrho_{\beta}(\widetilde{\Delta}_{\rm min})\,\langle p_{\rm LZ}\rangle_\gamma\nonumber\\
&\sim \tilde{v}^{1+\beta/2}.
\end{align}

\begin{figure*}[tb!]
\includegraphics[width=0.65\textwidth]{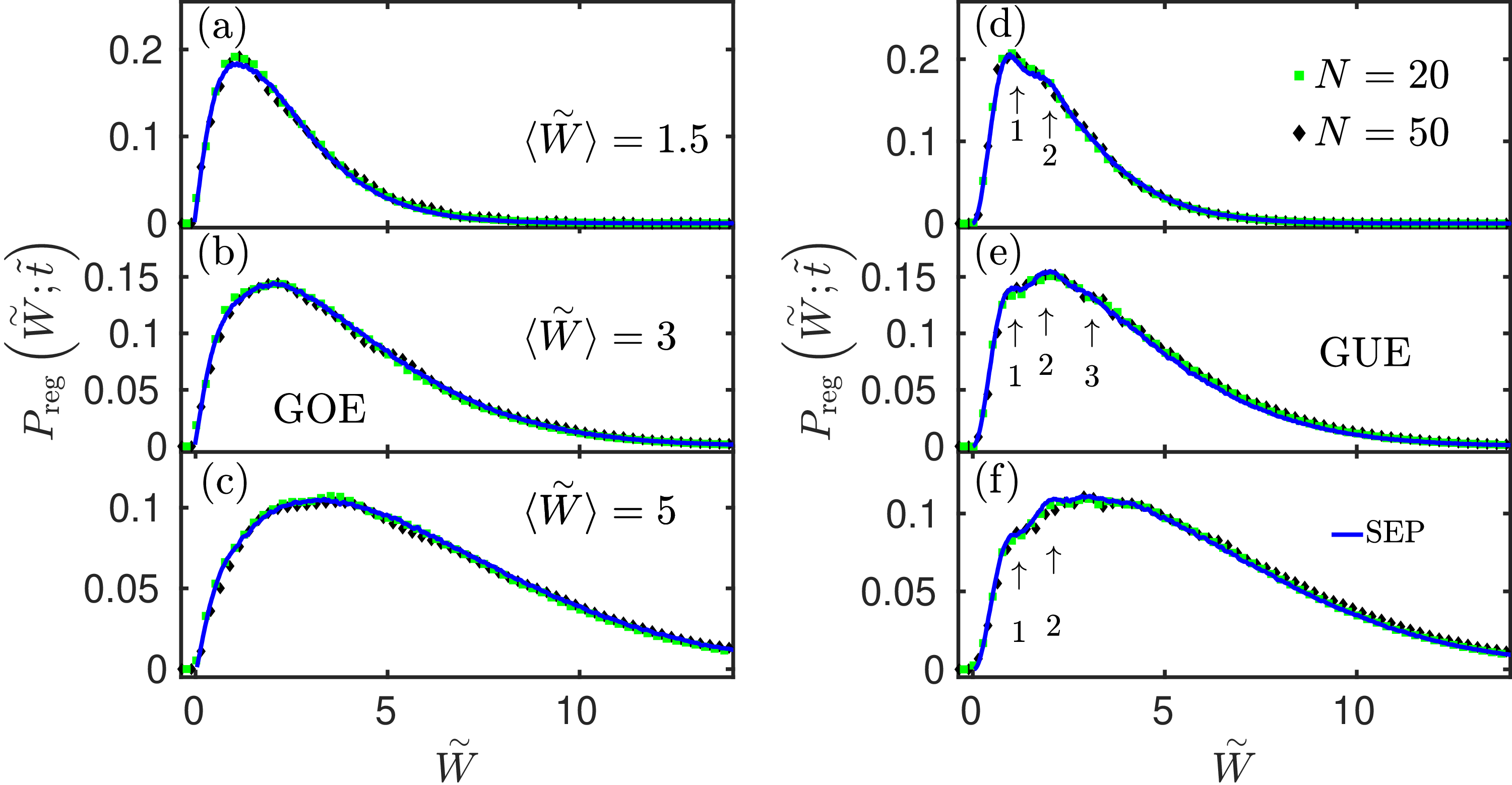}
\centering
\caption{Comparison of the regular part of the work statistics, $P_{\rm reg}(\Wtilde;\tilde t)$ for  (a)-(c) the orthogonal  (GOE) and (d)-(f) the unitary matrix ensembles (GUE), with 
$\Wtilde$ and $\tilde t$ referring to dimensionless work and time. 
Symbols: regular part for  different values of the dimensionless average work, $\langle \Wtilde \rangle = 1.5$, $3$, and $5$, and a dimensionless 
quench velocity, $\tilde{v} = 0.8$, for two different matrix sizes, $N = 20$ and $N = 50$. 
The distributions show universal behavior and collapse into a single curve independent of the system size, $N$. 
Arrows indicate the positions of the first neighboring levels at $\Wtilde = 1, 2, ...$.  
Solid lines: classical SEP simulations  provide an excellent  approximation for $\tilde{v} \lesssim 1$. In this limit, 
$P_{\rm reg}(\Wtilde;\tilde t)$ depends  only the  average dimensionless work, $\langle \Wtilde \rangle$, for both ensembles.
}
\label{fig:P(W)_small_work}
\end{figure*}

The universal scaling of the many-body diffusion constant, as extracted from our simulations, is plotted 
in Fig.~\ref{fig:D_v}. For time reversal symmetry breaking (unitary) systems, the  diffusion constant scales 
both at small and high velocities as ${\tilde v}^2$.
In the orthogonal case, however, a clear crossover is demonstrated  
between a Landau-Zener dominated   small velocity regime with $\widetilde D\sim {\tilde v}^{3/2}$, and a
high velocity fast quench regime with  $\widetilde D\sim \tilde v^{2}$ .

\section{Time evolution of single particle states}\label{sec:timeevol}
To solve Eq. (6) in the main text, it is practical to make a simple gauge transformation, 
and eliminate the phase dependence generated by the instantaneous eigenenergy, $\varepsilon_k(t)$,
$$
 \alpha_k(t) \equiv e^{-i \Theta_k(t)}  \tilde \alpha_k(t)\, 
$$
with the dynamical phase defined as 
\begin{equation}
\Theta_k(t)=\int_0^t dt^\prime\varepsilon_k(t^\prime).
\label{eq:dyn_phase}
\end{equation}
Importantly,  dynamical phases cancel  in the expression of the occupation numbers, 
$f_k(t)$, as well in the expression of the work and its generating function. 
Therefore, in all these equations,  we can  just replace $ \alpha_k(t)$ by $ \tilde \alpha_k(t)$, 
obeying the simpler differential equation, 
\begin{equation}
i \,\dot{\tilde \alpha}_k(t)=   \sum_{l} A^{kl}(t)\,\tilde\alpha_l(t)\,.
\label{eq:schrodinger2}
\end{equation}
In practical calculations,  it is useful 
to avoid numerical differentiation, and  calculate  the Berry connection as 
\begin{equation}
\label{eq:berry}
A^{kl}(t)=- i\,\langle \etavec^k_t|\partial_t \etavec^l_t\rangle= i\,\dfrac{\etavec^k_t \cdot \partial_t{\cal H}(t)\,{\etavec}^l_t}{\varepsilon_k(t)-\varepsilon_l(t)}.  
\end{equation}

In our numerics, we made use of the fourth order Runge-Kutta method to solve the single particle time-dependent Schr\"odinger equation.
Dynamical phases  in Eq.~\eqref{eq:dyn_phase}  were determined  numerically using  the Simpson-formula.  Since the phases of the instantaneous states often make  jumps,  we enforced  the choice ${A_{kk} = - i\;{\etavec}^k_t\cdot \partial_t {\etavec}^k_t=0}$
by requiring that  the overlaps of two consecutive eigenstates remain close to 1, $\langle \etavec^k_t |\etavec^k_{t+\delta t}\rangle \approx 1$.

\section{Fingerprints of level repulsion and deviations from universality}\label{sec:universality}

\begin{figure}[b!]
\includegraphics[width=1\columnwidth]{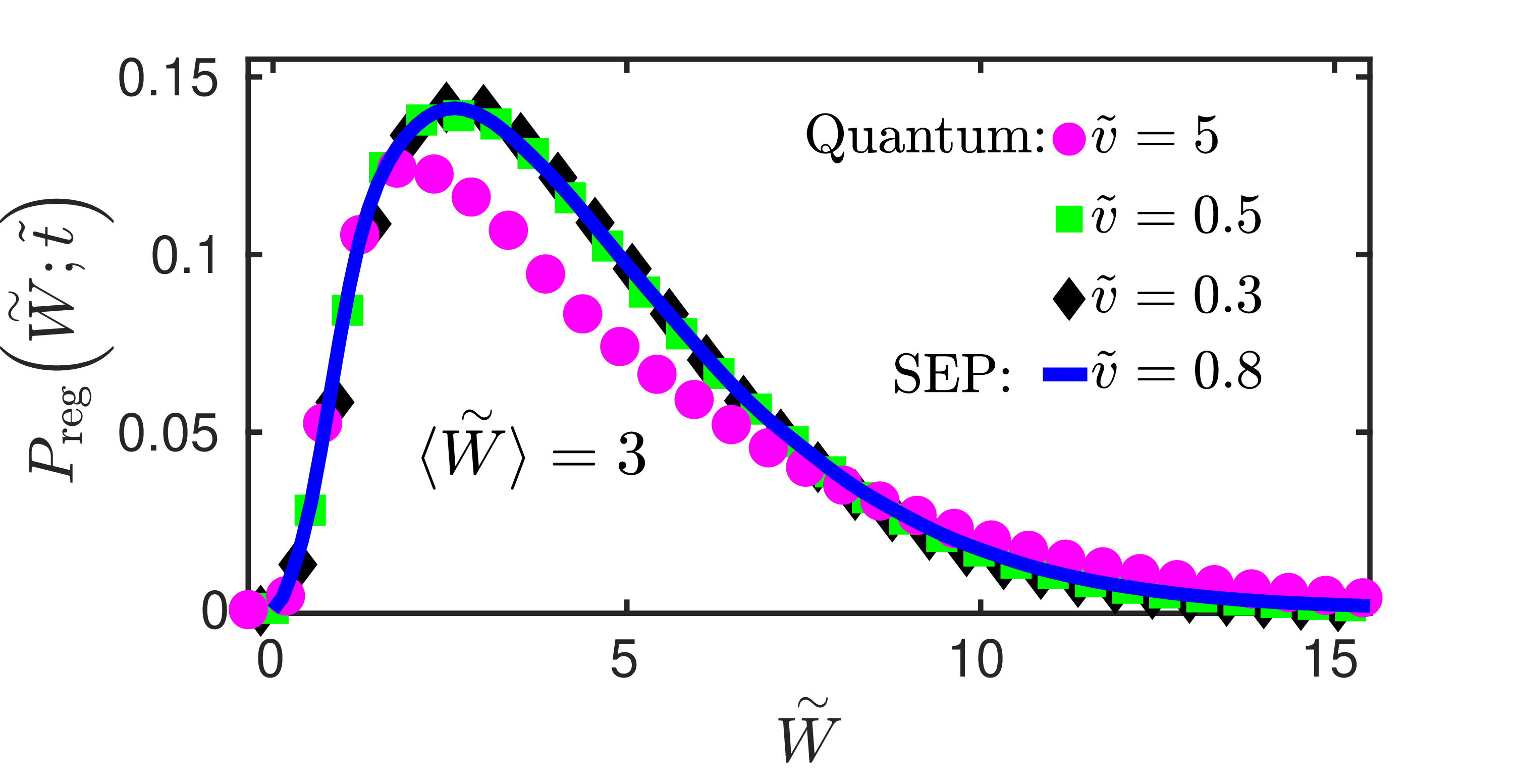}
\caption{
Universality of the distribution $P_{\rm reg}(\Wtilde;\tilde t)$  in the limit $\tilde{v} \lesssim 1$ for the orthogonal ensemble. 
Symbols: quantum results for $N = 20$ levels and  $\tilde{v} = 0.3$,  $\tilde{v} = 0.5$, and $\tilde{v} = 5$. 
Solid line: classical SEP  result for $\tilde{v} = 0.8$. While all $\tilde v\lesssim 1$ results fall on top of each other for the same amount of average work, 
for $\tilde{v} = 5$, deviations appear due to transitions between non nearest neighbor levels. }
\label{fig:universality}
\end{figure}

As discussed in the main text, the distribution of the work is universal, i.e., it is independent from the 
system size $N$ as well as the Fermi energy; it depends solely on the dimensionless time, 
the dimensionless velocity, and the symmetry class of the random Hamiltonians. 
This is illustrated in Fig.~\ref{fig:P(W)_small_work}, where we have fixed the  
dimensionless quench velocity to $\tilde{v} = 0.8$, and the dimensionless times such that they correspond 
to an injected dimensionless work $\langle \Wtilde \rangle = 1.5$, $3$, and $5$ in both symmetry classes. 

For relatively small injected internal energies, $W\lesssim 5\;\Deltabar $, the discreteness of the energy 
levels  becomes apparent, and fingerprints of  level repulsion can be observed. 
First,  $P(\Wtilde)$ vanishes continuously at zero (within  SEP as $\Wtilde{}\!\!^\beta$), since the first empty level is repelled from the last occupied level. 
Moreover, additional wiggles appear in $P(\Wtilde)$, reflecting the positions of first, second and third neighbors. These features are more pronounced  in the Gaussian unitary ensemble, where level repulsions are stronger, 
and are expected to become even more pronounced for the symplectic ensemble, not studied here. The observed 
distributions are, however, independent of the matrix size, $N$, as stated above.

Remarkably, even for $\tilde{v} = 0.8$,  SEP simulations (continuous lines in Fig.~\ref{fig:P(W)_small_work}) 
provide  an almost perfect description of the full quantum results. For larger velocities, however, 
deviations occur. This is demonstrated in Fig.~\ref{fig:universality}. Clearly, for 
$\tilde v\gtrsim 1$ the role of Landau-Zener transitions is reduced, 
the SEP description loses its validity, and distributions depend 
 explicitly on the velocity $\tilde v$.


\begin{references}
\bibitem{footnote_J} For simplicity, we set the usual energy unit to $J=1$ here.
\bibitem{footnoteA}We fix the gauge of the vectors $ \etavec^m_t$ such that $A^{kk}(t)\equiv 0$ at any time.
\bibitem{footnoteB} We used typically $\sim 30000$ samples to generate the work statistics.
\bibitem{footnote_f_k} This quantity can be computed without the determinant formula as 
 $f_k(t)= \langle\langle\,\hat n_{k,t}\rangle\rangle_\text{RM} = \sum_{m=1}^{M}\langle|\alpha_k^m(t)|^2\rangle_\text{RM}$.
\bibitem{footnoteC}We have also set up a slightly more complicated classical Markovian process to model the dynamics. Here we performed a Monte Carlo averaging over the matrices $\mathcal{H}_1$ and $\mathcal{H}_2$; for each pair $\{\mathcal{H}_1,\mathcal{H}_2\}$ we again intialized the fermions at the $M$ lowest levels. We then looked at the spectrum of $\mathcal{H}(t)$, and each time we encountered a Landau-Zener crossing such that only one of the two colliding levels was occupied, the fermion hopped to the other level with the Landau-Zener transition probability calculated for the particular level crossing. We have checked that this process yielded results very similar to the simplified version presented in the main text.
\bibitem{footnoterho}More generally, the density of avoided level crossings between any two neighboring levels depends also on the energy of the colliding levels, $\varepsilon$. By concentrating on the middle of the spectrum, here we restrict our analysis to energy $\varepsilon=0$.
\bibitem{footnote_pi}Our convention differs from the one used in Ref. \onlinecite{wilkinson3}, where $\widetilde{\Delta}$ is measured in units $1/N\equiv\overline{\Delta}/\pi$, leading to a factor of $\pi$ difference in the density $\tilde{\rho}_\Delta$.
\end{references}
\end{document}